# Three-Dimensional Variable Slab-Selective Projection Acquisition Imaging

Jinil Park, Taehoon Shin, and Jang-Yeon Park

***Abstract*—** Three-dimensional (3D) projection acquisition (PA) imaging has recently gained attention because of its advantages, such as achievability of very short echo time, less sensitivity to motion, and undersampled acquisition of projections without sacrificing spatial resolution. However, larger subjects require a stronger Nyquist criterion and are more likely to be affected by outer-volume signals outside the field of view (FOV), which significantly degrades the image quality. Here, we proposed a variable slab-selective projection acquisition (VSS-PA) method to mitigate the Nyquist criterion and effectively suppress aliasing streak artifacts in 3D PA imaging. The proposed method involves maintaining the vertical orientation of the slab-selective gradient for frequency-selective spin excitation and the readout gradient for data acquisition. As VSS-PA can selectively excite spins only in the width of the desired FOV in the projection direction during data acquisition, the effective size of the scanned object that determines the Nyquist criterion can be reduced. Additionally, unwanted signals originating from outside the FOV (e.g., aliasing streak artifacts) can be effectively avoided. The mitigation of the Nyquist criterion owing to VSS-PA was theoretically described and confirmed through numerical simulations and phantom and human lung experiments. These experiments further showed that the aliasing streak artifacts were nearly suppressed.

***Index Terms*—** 3D Projection acquisition, Nyquist criterion, Slab-selective projection, Streak artifact, Ultrashort echo-time (UTE).

## I. INTRODUCTION

AMONG the various spatial encoding schemes available in magnetic resonance imaging (MRI), radial acquisition (RA) or projection acquisition (PA) was the earliest strategy proposed by Lauterbur in 1973 [1], [2]. However, early MRI scanners performed poorly in terms of $B_0$ field homogeneity and gradient field linearity, resulting in unacceptable blurring artifacts in PA imaging. Nevertheless, recent advances in MRI hardware technology have enabled a robust scan environment, leading to the resurgence of PA-MRI [3], [4].

The key advantage of PA is its robustness to motion artifacts compared to Cartesian acquisition. Because all data acquired from the PA pass through the k-space center, oversampled low spatial frequencies are averaged over the motion, resulting in relatively marginal image blurring. Motion-induced data inconsistency at high frequencies produces streaks that are more tolerable in terms of pattern and strength than distinct ghost artifacts during Cartesian acquisition [5]. Additionally, each radial spoke contains information about the object's motion, allowing for motion estimation and compensation, as demonstrated by PROPELLER (Periodically rotated overlapping parallel lines with enhanced reconstruction) imaging techniques [6]. Finally, when half-spoke acquisition is used with center-out sampling, an ultrashort echo-time (UTE) is achieved owing to the absence of a phase-encoding gradient [4]. UTE imaging can capture signals of short $T_2/T_2^*$ species and has been used for imaging lung and musculoskeletal systems [7], [8], [9], [10], [11].

The number of full- or half-radial spokes required to satisfy the Nyquist criterion is $2\pi$ or $4\pi$ times larger, respectively, than the phase-encoding steps in Cartesian acquisition for the same spatial resolution and field of view (FOV). PA using sub-Nyquist sampling may yield benign aliasing artifacts owing to the variable density sampling nature. However, for large subjects, the occurrence of unacceptable streak artifacts is likely and can only be resolved by acquiring more spokes [12]. Acquiring multiple radial views per repetition time (TR) is one option to increase the sampling density with a marginal scan time cost but may cause image artifacts when the multi-echo data are inconsistent owing to system errors [13]. Another challenge of PA-MRI for a large subject is the gradient non-linearity problem, which tends to be more pronounced in regions far from the isocenter. The peripheral regions of a large subject, distant from the isocenter, may experience signal distortion owing to gradient non-linearity and sometimes produce high signal intensity outside the FOV, leading to strong streak-artifact signals inside the FOV [14], [15]. These issues associated with a large subject becomes particularly critical in three-dimensional (3D) PA imaging where aliased signals occur in all three dimensions.

Spatially selective spin excitation is an effective method to minimize artifacts resulting from insufficient sampling and gradient non-linearity in 3D PA imaging because an imaging object can be spatially limited during each TR. 3D spatially selective spin excitation is ideal for suppressing signals outside

This work was supported by the National Research Foundation of Korea (NRF) grant funded by NRF-2021R1C1C2008365 and NRF-2020R1A2B5B02002676; in part by the Ministry of Science and ICT under Grant RS-2023-00244602. (Corresponding authors: Jang-Yeon Park.)

Jinil Park and Jang-Yeon Park are with the Department of Intelligent Precision Healthcare Convergence, Sungkyunkwan University, Suwon, Republic of Korea (e-mail: jinilpark@skku.edu; jyparu@skku.edu).

Taehoon Shin is with Division of Mechanical and Biomedical Engineering, and with Graduate Program in Smart Factory, Ewha Woman University, Seoul, Republic of Korea (e-mail: shinage@gmail.com).

Jang-Yeon Park is with the Department of Biomedical Engineering, Sungkyunkwan University, Suwon, Republic of Korea (e-mail: jyparu@skku.edu).

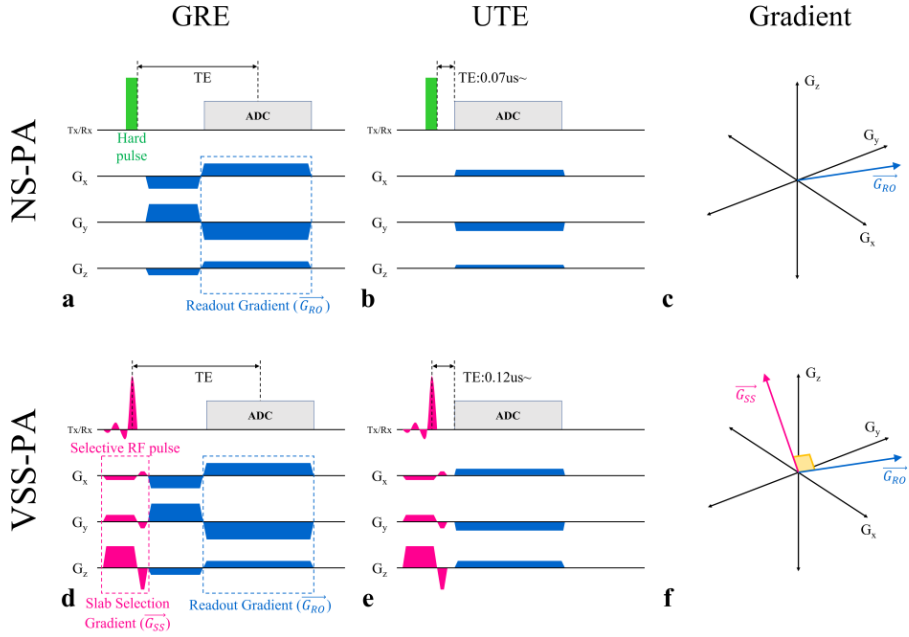

**Fig. 1** Sequence diagram of non-selective PA (NS-PA) and variable slab-selective PA (VSS-PA) shown as two versions, i.e., a gradient-echo (GRE) version with full echo and an ultrashort echo-time (UTE) version with half echo. NS-PA consists of a short hard RF pulse and a 3D readout gradient (a, b), and VSS-PA consists of a Shinnar Le Roux (SLR) RF pulse, a slab-selective gradient and a 3D readout gradient (d, e). Since the UTE version is acquired with half echo without dephasing gradients, ultrashort TE is available (b, e). Unlike NS-PA, which excites the entire object, VSS-PA uses slab-selective gradients and SLR pulses simultaneously to create spatially selective spin excitation (d, e). The magnitude of the slab-selective ($G_{SS}$) and readout ($G_{RO}$) gradients in the three axes ($G_x$, $G_y$, $G_z$) are expressed as 3D vectors. In NS-PA, only the readout gradient is present (c), whereas in VSS-PA, slab-selective and readout gradients always remain perpendicular (f).

the FOV in all three directions [16]. However, 3D frequency-selective radiofrequency (RF) pulses are long and rarely used in clinical practice. Instead, one-dimensional (1D) spatially selective spin excitation, often called slab-selective excitation in 3D MRI, is the most popular approach owing to its short RF pulse duration. However, when 1D spatially selective spin excitation is applied in the superior-to-inferior direction [10], it cannot remove aliasing artifacts occurring in the transverse plane, such as streak artifacts owing to the arms in lung imaging [17].

Another approach for reducing aliased signals is outer volume pre-saturation using magnetization preparation RF pulses [17]. Although aliasing artifacts can be further suppressed by this approach, the pre-saturation performance is often poor in the peripheral regions of subjects where the transmit RF error is large. Furthermore, segmented data acquisition is required after saturation preparation, with delay periods until the next preparation, which prolongs the total scan time. In addition to excitation-based approaches, post-processing techniques adaptively combine multichannel reception coils during image reconstruction [18], [19]. These techniques assign higher weights to RF channels containing regions of interest (ROI) with no artifacts, and lower weights to RF channels containing streak artifacts or minimal ROI. However, these methods tend to reduce the signal-to-noise ratio (SNR) and may even lose important SNR and may even lead to loss of important information if coil selection is suboptimal.

In this study, we proposed a data collection method that relaxes the Nyquist criterion to reduce streak artifacts in 3D PA-MRI. Unlike conventional slab-selective excitation applied in a fixed direction [10], the proposed method alters the direction of the slab-selective gradient to be perpendicular to the readout gradient, and is termed variable slab-selective PA (VSS-PA). This method excites only the desired FOV in each projection direction, reducing the effective size of the object and relaxing the Nyquist criterion. The performance of the VSS-PA was demonstrated using numerical simulations, phantom scans, and lung imaging of healthy subjects.

## II. METHOD

### A. Pulse Sequence

Fig. 1 illustrates the sequence diagram and vector representation of the readout and slab-selective gradients of non-selective PA (NS-PA) and VSS-PA MRI. Both gradient echo (GRE) and UTE imaging sequence versions were considered for each type of PA. The conventional PA sequence consists of a nonselective hard RF pulse, a predephasing gradient (only for the GRE version), and a readout gradient (Figs. 1(a) and 1(b)). The n$^{th}$ 3D readout gradient ($\vec{G}_{RO}$) for NS-PA can be written in spherical coordinates as

$$\vec{G}_{RO}(n) = \begin{bmatrix} G_{RO,x}(n) \\ G_{RO,y}(n) \\ G_{RO,z}(n) \end{bmatrix} = \begin{bmatrix} \cos\phi(n)\cos\theta(n) \\ \sin\phi(n)\cos\theta(n) \\ -\sin\theta(n) \end{bmatrix}, \quad (1)$$

where $\theta$ and $\phi$ are the polar and azimuthal angles, respectively. The 3D gradient vector is shown in Fig. 1(c). The UTE version enables ultrashort echo-time acquisition immediately after spin excitation, owing to the use of a very short hard RF pulse (Fig. 1(b)).

The proposed VSS-PA employs slab-selective RF and gradient pulses to spatially minimize the spin excitation in the area outside the FOV

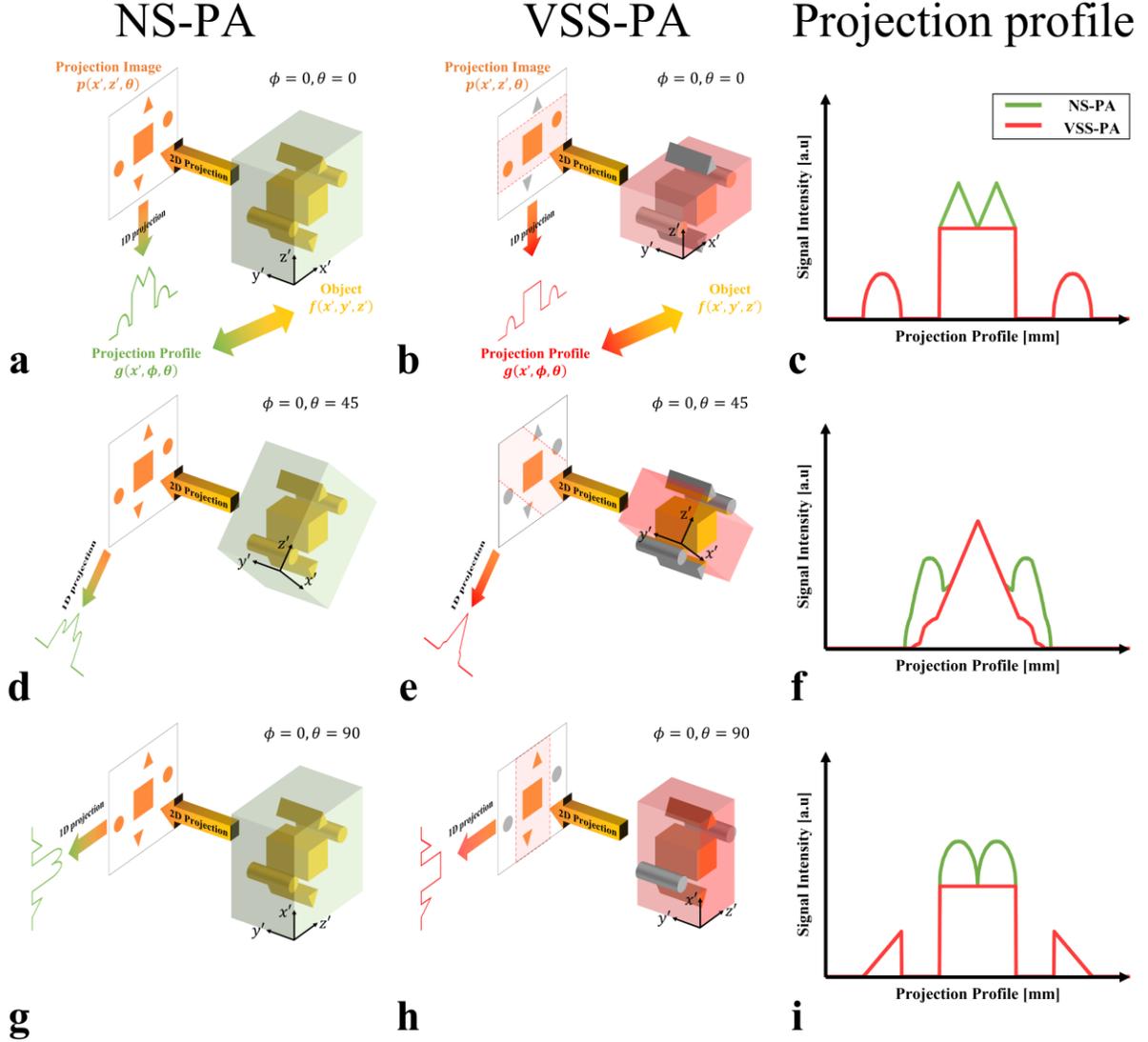

**Fig. 2** Comparison of acquisition schemes and projection profiles (c, f, i) of non-selective PA (NS-PA) (a, d, g) and variable slab-selective PA (VSS-PA) (b, e, h), describing the 3D Radon transform through two-step projections, i.e., 2D and 1D projections. The acquisition system is shown through three angle examples. NS-PA excites the entire subject (green box) using non-selective RF pulses, so all information is included in the 1D projection profile (a, d, g). Since VSS-PA performs slab-selective excitation (red box) of width $L$ in the $z'$ direction, the projection along the $z'$ direction contains only spatially selective information (b, e, h). The 1D profile obtained from NS-PA contains information of two triangular prisms and two cylinders other than the hexahedron, whereas VSS-PA does not include or minimize the two triangular prism and two cylinder information.

(Figs. 1(d) and 1(e)). The slab-selective gradient ($\vec{G_{SS}}$) is modified to be perpendicular to the direction of the readout gradient. The orthogonal relationship between $\vec{G_{RO}}$ and $\vec{G_{SS}}$ is illustrated in Fig. 1(f). The $\vec{G_{SS}}$ satisfying this orthogonal condition is defined as:

$$\vec{G_{SS}}(n) = \begin{bmatrix} G_{SS,x}(n) \\ G_{SS,y}(n) \\ G_{SS,z}(n) \end{bmatrix} = \begin{bmatrix} \cos\phi(n)\sin\theta(n) \\ \sin\phi(n)\sin\theta(n) \\ \cos\theta(n) \end{bmatrix} \quad (2)$$

Note that a minimum-phase Shinnar Le Roux (SLR) [20] design was used to generate a frequency-selective asymmetric RF pulse in the VSS-PA of both the GRE and UTE versions (Figs. 1(d) and 1(e)). The resulting echo time (TE) of VSS-PA was only 0.05 μs longer than that of NS-PA.

### B. Acquisition Scheme

According to the projection-slice theorem, the Fourier transform of a radial spoke acquired in k-space is equivalent to the projection of an object in the direction perpendicular to the radial spoke in the image domain, that is, the Radon transform of an object in that direction. Therefore, when the entire object $f(x, y, z)$ is spin-excited, the 3D NS-PA can be expressed as

$$g(x') = \int_{-\infty}^{\infty}\int_{-\infty}^{\infty} f(x', y', z') dy' dz', \quad (3)$$

where $g(x')$ is the resultant 1D projection profile along the $x'$ direction, and $(x', y', z')$ represents the rotating coordinate system defined by polar angle $\theta$ and azimuthal angle $\phi$ with a reference to the laboratory coordinate system $(x, y, z)$ as follows:

$$\begin{aligned} x' &= x\cos\phi\cos\theta - y\sin\phi + z\cos\phi\sin\theta \\ y' &= x\sin\phi\cos\theta + y\cos\phi + z\sin\phi\sin\theta \\ z' &= -x\sin\theta + z\cos\theta \end{aligned} \quad (4)$$

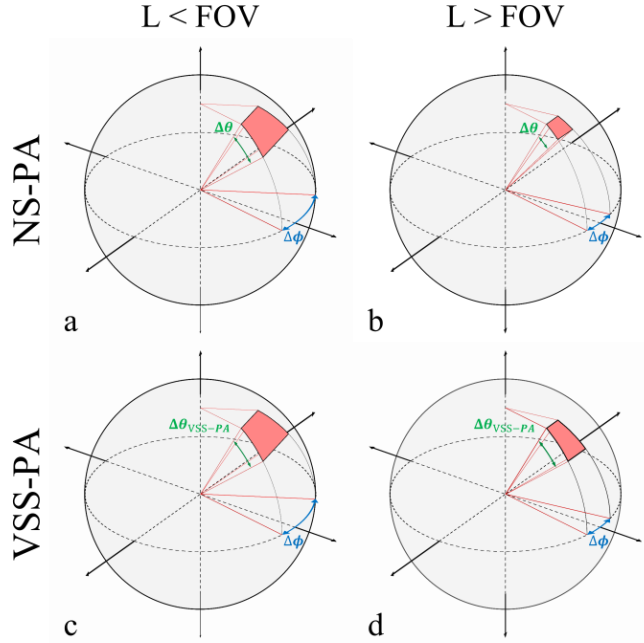

**Fig. 3** Visualization of the Nyquist criterion of non-selective PA (NS-PA) (a, b) and variable slab-selective PA (VSS-PA) (c, d) in k-space for 3D projection acquisition. When the size of an object is smaller than the FOV, the Nyquist criterion is the same for NS-PA (a) and VSS-PA (c). On the other hand, when the size of an object is larger than the FOV, Nyquist criterion of NS-PA requires more radial spokes or projections because Δϕ and Δθ decrease as the object size increases (b). Since VSS-PA uses slab-selective projection, Δϕ is the same as NS-PA, but Δθ is not affected by the object size (d). Therefore, when the size of an object increases, a relatively relaxed Nyquist criterion is required compared with NS-PA.

This 3D projection can be divided into two steps: a two-dimensional (2D) projection and a 1D projection, as shown in Figs. 2(a), 2(d), and 2(g). In the first step, a 2D projection image $p(x', z')$ is obtained by projecting the 3D object onto the $x'z'$-plane along the $y'$ direction; that is,

$$p(x', z') = \int_{-\infty}^{\infty} f(x', y', z') dy' \quad (5)$$

In the second step, a 1D projection profile is acquired through a line projection of the 2D projection image along the $z'$ direction.

$$g(x') = \int_{-\infty}^{\infty} p(x', z') dz' = \int_{-\infty}^{\infty}\int_{-\infty}^{\infty} f(x', y', z') dy' dz' \quad (6)$$

The first column of Fig. 2 illustrates the Radon transform of the simulated 3D object, which consists of a hexahedron-shaped object of interest surrounded by four outer-volume objects, that is, two triangular prisms and two cylinders. Owing to non-selective spin excitation, all the objects are encoded in both 2D projection images ($p(x', z')$ in (5)) and 1D projection profiles ($g(x')$ in (6)). Therefore, the NS-PA data acquired in all orientations were contaminated by unwanted signals from the outer-volume triangular prisms and cylinders (Figs. 2(a), 2(d), and 2(g)).

However, the proposed VSS-PA significantly reduces the unwanted signal contamination from outer-volume objects by performing slab-selective spin excitation using the slab-selective gradient $\overrightarrow{G_{SS}}$ in varying $z'$ directions orthogonal to the readout gradient $\overrightarrow{G_{RO}}$. In this case, the object is slab-selected in the $z'$ direction, and $f(x', y', z')$ is changed to $f_{ss,z'}(x', y', z')$ in (3), where the subscript $ss$ denotes slab-selected:

$$g(x') = \int_{-\infty}^{\infty}\int_{-\infty}^{\infty} f_{ss,z'}(x', y', z') dy' dz' \quad (7)$$

Therefore, for the VSS-PA, the 2D projection (4) is slab-selected in the $z'$ direction, and the subsequent 1D projection (5) along the $z'$ direction suppresses the unwanted outer-volume signals.

Figs. 2(b), 2(e), and 2(h) illustrate the two steps of 3D projection (i.e., 2D projection followed by 1D projection) for VSS-PA at three representative projection orientations. For instance, when the projection orientation is $\phi = 0°$ and $\theta = 0°$ (Fig. 2(b)), unlike in NS-PA, the signals generated by the two triangular prisms are excluded through 2D projection followed by 1D projection. Similarly, when the projection orientation is $\phi = 0°$ and $\theta = 45°$ (Fig. 2(e)), the projection of NS-PA includes signals from both the cylinders and the triangular prisms, whereas the projection of VSS-PA includes signals from some of these objects and excludes from others. Finally, when the projection orientation is $\phi = 0°$ and $\theta = 90°$ (Fig. 2h), signals from the cylinders are excluded in the projections of VSS-PA.

### C. Nyquist Criterion

VSS-PAs has advantages over NS-PA in terms of the Nyquist criterion. They effectively minimize unwanted outer volume signals and also reduce the number of radial spokes that satisfy the Nyquist criterion, thereby reducing the scan time while maintaining spatial resolution. In 3D PA-MRI, the minimum condition between the endpoints of neighboring radial spokes needed to satisfy the Nyquist criterion is expressed through solid-angle sampling as follows:

$$(k_{max})^2 \Delta\Omega = \frac{1}{L^2}, \quad (8)$$

where $k_{max}$ is the maximum value of the 3D k-space, $\Delta\Omega$ is the solid-angle sampling interval, and $L$ is the largest dimension of an object. The number of full-echo spokes ($N_s$) required to fill the surface of the k-space sphere is calculated as

$$N_s = \frac{2\pi}{\Delta\Omega} = 2\pi(k_{max}L)^2 \quad (9)$$

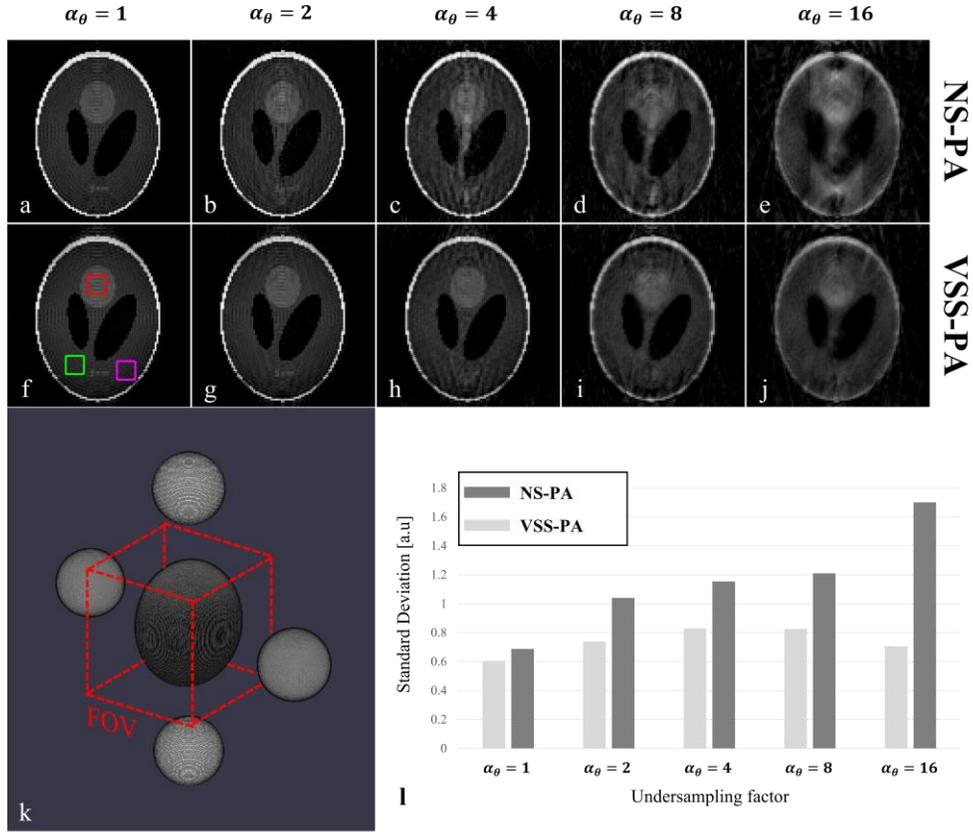

**Fig. 4** Phantom simulation results due to undersampling of non-selective PA (NS-PA) and variable slab-selective PA (VSS-PA). The 3D phantom used for simulation is shown in (k). The 3D Shepp-Logan phantom is surrounded by four spherical phantoms. The red dotted cube represents the simulated FOV. In the case of NS-PA, as the undersampling factor increases, streak artifacts become more severe and image quality deteriorates (a-c). On the other hand, streak artifacts appear as the undersampling factor increases, but image quality is relatively maintained compared to NS-PA. (f-j). The degree of streak artifacts was assessed in the three ROIs (red, green, and magenta boxes) (f) by measuring standard deviations in the reconstructed images (l).

According to (9), for NS-PA, which excites the entire object, the number of radial spokes required to satisfy the Nyquist criterion increases quadratically with the object size.

In contrast, VSS-PA has a different Nyquist criterion owing to slab-selective excitation along the $z'$ direction at every projection orientation. Considering the Nyquist criterion in the $\theta$ and $\phi$ directions, separately, $\Delta\Omega$ in (8) can be expressed as

$$\Delta\Omega = \sin\theta \cdot \Delta\theta \cdot \Delta\phi \quad (10)$$

Here, the minimum azimuthal sampling interval $\Delta\phi$ to satisfy the Nyquist criterion is given by

$$\Delta\phi = \frac{1}{k_{max} \cdot L \cdot \sin\theta} \quad (11)$$

which depends simply on the object size $L$ as in NS-PA. On the other hand, due to the spatially selected spin excitation of slab thickness $L_{slab}$ by applying $\overrightarrow{G_{SS}}$ perpendicular to $\overrightarrow{G_{RO}}$, the minimum polar sampling interval $\Delta\theta$ to satisfy the Nyquist criterion in VSS-PA is given by

$$\Delta\theta_{VSS-PA} = \frac{1}{k_{max} \cdot L_{slab}} \quad (12)$$

Note that $\Delta\theta_{VSS-PA}$ is independent of the object size because $k_{max}$ and $L_{slab}$ are associated with experimental parameters that can be controlled.

Therefore, considering both $\Delta\phi$ and $\Delta\theta_{VSS-PA}$, the number of radial spokes to satisfy the Nyquist criterion for VSS-PA can be written as

$$N_{s,VSS-PA} = \frac{2\pi}{\sin\theta \cdot \Delta\phi \cdot \Delta\theta_{VSS-PA}} = 2\pi(k_{max})^2(L \cdot L_{slab}) \quad (13)$$

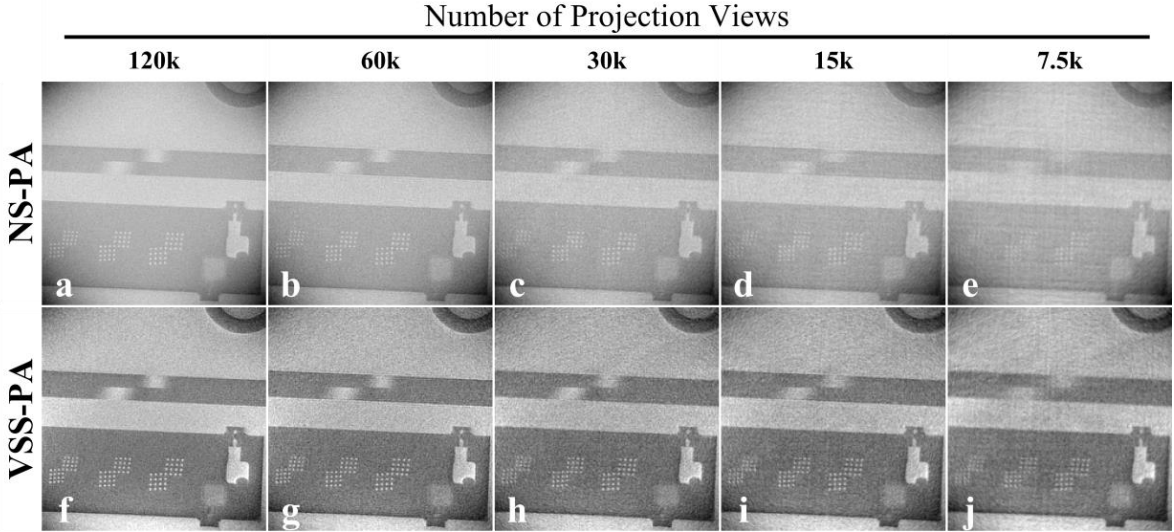

**Fig. 5** Enlarged images of a resolution phantom including resolution grids acquired from non-selective PA (NS-PA) (a–e) and variable slab-selective PA (VSS-PA) (f–j) to validate numerical phantom simulations. The streak artifacts increase in both NS-PA and VSS-PA images as the number of projections decreases. However, VSS-PA maintains image quality relatively well compared to NS-PA (f-j). In particular, In the most undersampled case, when the number of projections is 7.5k, the internal structure is barely discernible in NS-PA, but still discernible in VSS-PA.

Compared with NS-PA (9), the required number of radial spokes was reduced by a factor of $L/L_{\text{slab}}$.

Fig. 3 shows the solid angle formed by $\Delta\phi$ and $\Delta\theta$ required to satisfy the Nyquist criterion for NS-PA and VSS-PA, respectively, depending on the object size ($L$). When the object size is smaller than the FOV, both nonselective PA and VSS-PA require the same size solid angle (Figs. 3(a), 3(c)). However, if the object size is larger than the FOV, the polar angle interval required by NS-PA and VSS-PA, i.e., $\Delta\theta$ and $\Delta\theta_{VSS-PA}$, differs. For NS-PA, Nyquist criterions for $\theta$ and $\phi$ are both affected by the object size, so the sampling interval is reduced in both $\theta$ and $\phi$ directions (Fig. 3(b)). However, for VSS-PA, the Nyquist criterion in the $\theta$ direction remains unaffected by the object size owing to the slab-selective projection. Therefore, the sampling interval does not have to decrease in the $\theta$ direction (Fig. 3(d)), thereby enabling the sampling of the entire k-space with fewer radial spokes than NS-PA.

For example, if an object of $L$ = 200 mm is scanned with FOV = $100 \times 100 \times 100$ mm³ and matrix size = $100 \times 100 \times 100$, the number of radial spokes necessary to satisfy the Nyquist criterion is 62,831 and 31,415 for the PA and VSS-PA, respectively.

## III. SIMULATION AND EXPERIMENTS

The NS-PA and VSS-PA sequences were implemented and tested on a clinical 3T scanner (MAGNETOM Prisma; SIEMENS, Erlangen, Germany) with a maximum gradient amplitude of 80 mT/m and a maximum slew rate of 200 mT/m/ms. The GRE and UTE versions of the NS-PA and VSS-PA sequences were implemented in the IDEA pulse programming environment (SIEMENS, Erlangen, Germany) for the phantom and human lung experiments, respectively. To implement the UTE version of the VSS-PA, an SLR pulse was used to reduce TE by minimizing the phase dispersion of the transverse magnetization resulting from the slab-select gradient. For offline image reconstruction, gridding reconstruction was performed using a Kaiser-Bessel interpolation kernel with a diameter of 4 in MATLAB (MathWorks Inc. software (Natick, MA) [21].

### A. Numerical Simulation

Numerical simulations were performed by changing the undersampling factor to verify the Nyquist criterion according to the object size in the NS-PA and VSS-PA schemes. The number of radial spokes that satisfy the Nyquist criterion of the NS-PA ($N_s$ in (9)) was set to 62,831. These fully sampled k-space data were undersampled in the $\theta$ direction, with an undersampling factor ranging from 1 to 16 in increments of 2. A 3D image containing four spherical objects around a 3D Shepp-Logan phantom was simulated, and the FOV was set to tightly enclose only the Shepp-Logan phantom (Fig. 4(k)). In the VSS-PA, slab-selective excitation was simulated by multiplying the cuboid function of the FOV height in each projection direction. For NS-PA and VSS-PA, 3D projection acquisition was performed in a manner identical to that in the experiment using the inverse gridding method. Data were acquired in the form of a full echo to create the same environment as the GRE version of NS-PA. The other simulation parameters were as follows: $L = 200 \times 200 \times 200$ mm³, FOV = $100 \times 100 \times 100$ mm³, $L_{Slab}$ =100 mm, and the acquisition matrix = $100 \times 100 \times 100$.

### B. Phantom Experiment

A phantom experiment was performed to confirm the numerical simulation results. The scanned phantom consisted of an ACR phantom surrounded by a cylindrical object and a spherical object, with the FOV set to tightly include the ACR phantom. Both 18-channel chest and 12-channel spinal coils were used for signal reception. The number of radial spokes was set to 120,000 and undersampled in the $\theta$ direction, with an undersampling factor ranging from 1 to 8 in increments of 2. The scan parameters were as follows: TR = 6 ms, TE (NS-PA/VSS-PA) = 0.07/0.12 ms, FOV= $100 \times 100 \times 100$ mm³, flip angle = 2°, bandwidth per pixel = 770Hz, spatial resolution = $0.5 \times 0.5 \times 0.5$ mm³, matrix size = $200 \times 200 \times 200$.

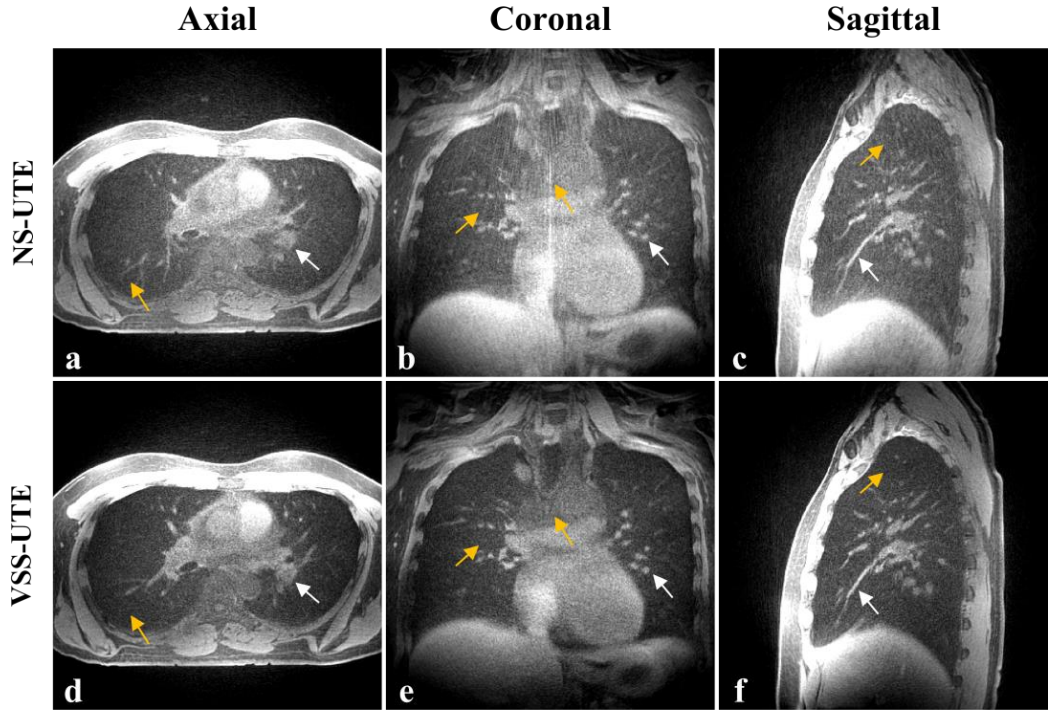

**Fig. 6** Comparison of axial, coronal and sagittal slices of human lungs between non-selective UTE (NS-UTE) (a-c) and variable slab-selective UTE (VSS-UTE) (d-f). On NS-UTE images, streak artifacts originating from the neck and abdomen appear inside the lungs, whereas VSS-UTE effectively suppresses streak artifacts in axial and coronal slices due to slab-selective projection.

A square pulse with a length of 20 μs was used for the NS-PA; A 150-μs-long SLR pulse designed with a bandwidth of 6.4kHz was used for the VSS-PA.

### C. Human Lung Experiment

*In vivo* human lung imaging was performed on one healthy volunteer during free breathing. The lungs had a very short $T_2$ and $T_2^*$ owing to the low proton density and susceptibility difference between air and tissue, requiring imaging with a very short TE to minimize signal loss. Therefore, the UTE version was used for both 3D NS-PA and 3D VSS-PA (Fig. 1(b) and 1(e)). Fat appears bright in UTE images because of the image contrast determined by a combination of proton-density and $T_1$ weighting. Spectrally selective fat saturation RF pulses were used in every 10 projections to mitigate streak artifacts originating from fat signals in the undersampled PA data.

To minimize the respiratory motion, retrospective respiratory gating was performed using the self-navigation method proposed in [22] with a respiratory gating efficiency of 45%. Self-navigation signals that track respiratory movements may be corrupted owing to the intermittent application of the fat-saturation preparation. To address this issue, self-navigation signals were acquired at the end of the segmented data acquisition.

Both 18-channel chest and 16-channel spine coils were used for signal reception. The FOV was positioned to avoid including the arms and neck of the volunteer to prevent streak artifacts. Scan parameters were as follows: TR = 3 ms, TE(UTE/VSS-UTE) = 0.07/0.14 ms, FOV= 360×360×360 mm³, flip angle = 5°, bandwidth per pixel = 730Hz, spatial resolution = 0.81 × 0.81 × 0.81 mm³, matrix size = 440 × 440 × 440, and the number of projections = 120,000.

## IV. RESULT

### A. Numerical Simulation

Fig. 4 shows the simulated images (Fig. 4(a)-(j)) acquired using NS-PA and VSS-PA with varying undersampling factors indicated by $\alpha$. When $\alpha = 1$, which met the Nyquist criterion, both the NS-PA and VSS-PA images showed the Shepp-Logan phantom without streak artifacts (Fig. 4(a), 4(f)).

As the $\alpha$ value increased (i.e., the number of radial spokes decreases), streak artifacts appeared stronger in both the NS-PA and VSS-PA simulations. However, it was noticeable that NS-PA was more severely affected by the undersampling of the radial spokes, as the Nyquist criterion in terms of $\Delta\theta$ was stronger than that of VSS-PA (Fig. 4(a)-(e)). For example, when $\alpha = 16$, the streak artifacts were severe in the NS-PA images, making it difficult to recognize the phantom structure. However, the image quality of VSS-PA is relatively well maintained even at large values of $\alpha$ due to the slab-selective projection (Fig. 4(f)-(j)).

Fig. 4(l) shows bar graphs comparing the standard deviation of pixel values in the three ROIs, where the signal intensities should ideally be uniform. While the standard deviation increased with increasing $\alpha$ for VSS-PA owing to the increasing aliasing effects, VSS-PA yielded an overall lower standard deviation than NS-PA for all values of $\alpha$ (Fig. 4(l)).

### B. Phantom Experiment

Fig. 5 shows the resolution grid images of the 3D ACR phantom acquired using NS-PA and VSS-PA. Although both NS-PA and VSS-PA degraded the image quality as the number of projections decreased, streak artifacts were less evident in VSS-PA than in NS-PA. In particular, when the number of projections was only 7.5k, NS-PA could not identify each resolution grid, whereas VSS-PA could. This is because the projection data in VSS-PA are less contaminated by the outer-volume signal.

### C. Human Lung Experiment

56% of all the acquired projections were selected through retrospective respiratory gating and used for image reconstruction. The proportion of selected projections was set equally for nonselective UTE (NS-UTE) and variable slab-selective UTE (VSS-UTE) for a fair comparison. Fig. 6 shows representative axial (Fig. 6(a), 6(d)), coronal (Fig. 6(b), 6(e)), and sagittal (Fig. 6(c), 6(f)) slices of the healthy volunteer acquired using NS-UTE and VSS-UTE. Streak artifacts originating from the neck, arms, and abdomen were conspicuous in the NS-UTE images (yellow arrows). In the VSS-UTE images, these artifacts were barely visible owing to the slab-selective PA in each radial spokes, allowing for better identification of internal vessel structures (white arrows).

## V. Discussion

The proposed 3D VSS-PA imaging technique aims to reduce the number of projections that satisfy the Nyquist criterion and minimize the inclusion of unwanted signal sources outside the FOV. The proposed method acquires data by using slab-selective excitation in all projection acquisitions, and setting the readout gradient perpendicular to the slab-selective gradient (Fig. 1(f)). In this case, the effective object size relative to the Nyquist criterion is reduced because of the slab-selective projection in all projection directions, reducing the likelihood of streak artifacts in undersampled projection data.

The effectiveness of the proposed method depends on the size of the object, which is determined by the RF coil sensitivity relative to the FOV. When computing the Nyquist criterion with it, VSS-PA and NS-PA show no difference unless the object size exceeds the FOV (Fig. 3(a), 3(c)). However, VSS-PA has a clear advantage over NS-PA when the FOV is smaller than the object size. This scenario commonly occurs in practice, because it is not easy to completely satisfy the Nyquist criterion in a limited scan time.

Although spin excitation using the proposed method is slab-selective, aliasing effects cannot be completely excluded in all projection directions. For example, aliasing in the azimuthal direction (ϕ) is affected by the object size. However, in VSS-PA, aliasing in the polar direction (θ) can be minimized using (12). As demonstrated in the simulation (Fig. 4) and phantom experiment (Fig. 5), the proposed VSS-PA method maintained the image quality relatively well when undersampled in the polar direction. The aliasing effect also varies depending on the shape of the object and the trajectory settings of the radial spokes.

Undersampling of the radial spokes is inevitable for 3D PA imaging of organs sensitive to respiratory motion, such as the lungs, where retrospective respiratory gating is necessary. As discussed earlier, VSS-PA benefits from a relaxed Nyquist criterion over NS-PA when the object size, as determined by the RF coil sensitivity, exceeds the FOV. Therefore, this feature of VSS-PA is advantageous for undersampled projections in 3D PA imaging and will be even more useful when combined with undersampled imaging and reconstruction techniques such as parallel imaging [23], [24], [25], compressed sensing [26] and artificial intelligence [27].

Although the proposed VSS-PA method is based on 3D radial k-space sampling, it can be applied to other types of k-space sampling methods. For example, if data are acquired by spiral (or radial) acquisition in a 2D k-space plane perpendicular to the slab-selective gradient direction, the 2D fast Fourier transform of these data produces a 2D projection image unaffected by outside the slab-selective bandwidth. The acquired 2D k-space plane can be rotated about an arbitrary axis to fill 3D k-space space [28]. This acquisition strategy may have similar benefits as VSS-PA because it can also relax the Nyquist criterion.

In this study, we introduced a 3D variable VSS-PA MRI technique that acquires spatially selective signals for every projection acquisition. This was achieved by applying a slab-selective orthogonal gradient to the readout gradient (Fig. 1(f)). The proposed VSS-PA method not only relaxed the Nyquist criterion but also effectively suppressed unwanted signals such as aliasing streak artifacts originating outside the FOV.